\documentclass[12pt]{article}

\usepackage[letterpaper,top=2cm,bottom=2cm,left=2cm,right=2cm,marginparwidth=1.75cm]{geometry}
\usepackage{amsmath,amssymb}
\usepackage{graphicx}
\usepackage[colorlinks=true, allcolors=blue]{hyperref}
\usepackage{authblk}
\usepackage{tabularray}
\usepackage{bbm}
\usepackage{multirow}
\usepackage{cite}
\UseTblrLibrary{booktabs}

\def\mathswitchr#1{\relax\ifmmode{\mathrm{#1}}\else$\mathrm{#1}$\fi}
\def\mathswitch#1{\relax\ifmmode#1\else$#1$\fi}
\newcommand{\brc}[1]{\left(#1\right)}

\newcommand{\corr}{\mathswitchr{Corr}}
\newcommand{\bbeta}{\mathswitch{\boldsymbol{\beta}}}
\newcommand{\bmu}{\mathswitch{\boldsymbol{\mu}}}
\newcommand{\bA}{\mathswitch{\boldsymbol{A}}}
\newcommand{\bC}{\mathswitch{\boldsymbol{C}}}
\newcommand{\bI}{\mathswitch{\boldsymbol{I}}}
\newcommand{\bR}{\mathswitch{\boldsymbol{R}}}
\newcommand{\bU}{\mathswitch{\boldsymbol{U}}}
\newcommand{\bV}{\mathswitch{\boldsymbol{V}}}
\newcommand{\bZ}{\mathswitch{\boldsymbol{Z}}}

\newcommand{\declare}[2]{\vspace{2em}\noindent{\fontsize{14}{14}\selectfont\textbf{#1}}{%
\par\vspace{3pt}{\fontsize{12}{14}\selectfont #2}\par}}

\usepackage{listings}
\title{\textbf{\Large Homogeneity Test of Proportions for Combined Unilateral and Bilateral Data via GEE and MLE Approaches}}
\author{Jia Zhou \footnote{\href{mailto:jiazhou@buffalo.edu}{jiazhou@buffalo.edu}} }
\author{Chang-Xing Ma \footnote{\href{mailto:cxma@buffalo.edu}{cxma@buffalo.edu}}}
\affil{Department of Biostatistics, University at Buffalo, Buffalo, NY 14214, USA}
\date{}

\begin{document}
\maketitle

\begin{abstract}
\fontsize{12pt}{14pt}\selectfont

In clinical trials involving paired organs such as eyes, ears, and kidneys, binary outcomes may be collected bilaterally or unilaterally. In such combined datasets, bilateral outcomes exhibit intra-subject correlation, while unilateral outcomes are assumed independent.   
We investigate the generalized Estimating Equations (GEE) approach for testing homogeneity of proportions across multiple groups for the combined unilateral and bilateral data, and compare it with three likelihood-based statistics (likelihood ratio, Wald-type, and score) under Rosner’s constant $R$ model and Donner’s equal correlation $\rho$ model. Monte Carlo simulations evaluate empirical type I error and power under varied sample sizes and parameter settings. The GEE and score tests show superior type I error control, outperforming likelihood ratio and Wald-type tests. Applications to two real datasets in otolaryngologic and ophthalmologic studies illustrate the methods. We recommend the GEE and score tests for homogeneity testing, and suggest GEE for more complex models with covariates, while favoring the score statistic for small sample exact tests due to its computational efficiency. 

\end{abstract}

\section{Introduction}
\label{sec:intro}
In comparative clinical trials, binary bilateral observations often consist of paired organ (e.g., eyes, ears and kidneys) records in dichotomous form, with ``1'' denoting a cured or affected status and ``0'' otherwise. In practice, such datasets often comprise a mixture of bilateral and unilateral observations, since some subjects contribute paired organ records while others only a single record, for reasons such as prior surgical removal, congenital absence, or missing measurement for one organ. It is therefore essential to analyze the combined bilateral and unilateral data together, rather than discarding unilateral observations, to avoid loss of valuable information and potential bias in inference. Furthermore, appropriate modeling of the intra-subject correlation inherent in paired bilateral outcomes is necessary to accurately capture the dependence structure and ensure valid statistical conclusions. 
Rosner~\cite{Rosner_1982} first proposed the ``constant $R$ model" to test whether the proportion of affected eyes is equal across all groups, assuming equal dependence between the two eyes of the same individual. Dallal~\cite{Dallal_1988} critiqued Rosner's model for its poor fit when the binary status (e.g., cured or affected) is present bilaterally yet shows substantially different prevalences across groups. To address this, he proposed a model where the conditional probability of a response in one organ, given a response in the other, is a constant. Donner~\cite{Donner1989rhoModel} introduced an alternative approach assuming a constant intra-subject correlation across all individuals, which Thompson~\cite{Thompson_1993} subsequently demonstrated to be robust through simulation studies.  

Methods based on maximum likelihood estimates (MLEs) and likelihood-based tests under the foregoing models have been developed in numerous subsequent studies. For example, Tang \textit{et al.}~\cite{Tang_2008} proposed an asymptotic method for testing the equality of proportions between two groups under Rosner's model for correlated binary data. Ma \textit{et al.}~\cite{Ma_2013R} and Ma and Liu~\cite{ma2017rho} investigated three tests (likelihood ratio, Wald-type and score) for testing homogeneity among $g\ge2$ groups under Rosner's and Donner's models, respectively. Ma and K. Wang~\cite{ma2021testing} and Ma and H. Wang~\cite{ma2022testing} further examined several test procedures to test homogeneity of general $g\ge2$ proportions for combined unilateral and bilateral data under Rosner’s and Donner's models, respectively. 

Alternatively, regression models for correlated paired organ data can be used to perform hypothesis testing~\cite{ying2017tutorial,ying2018tutorial}, with the generalized linear mixed model (GLMM)~\cite{breslow1993approximate} and generalized estimating equations (GEE)~\cite{liang1986} being the most common approaches. Both methods are implemented in standard statistical software such as \texttt{SAS}, \texttt{R} and \texttt{Stata}. 
The GEE method, introduced by Liang and Zeger for longitudinal data~\cite{liang1986} and later extended to clustered data~\cite{liang1993regression}, generalizes the generalized liner model (GLM) framework to account for within-cluster dependence via estimating equations. 
It has been widely adopted for longitudinal and clustered data, and is often prefered over GLMM because it yields consistent parameter estimates and robust (co)variance estimates even when the ``working” correlation structure is misspecified.

Bilateral observations from the same subject can be regarded as repeated measurements on that individual, making the GEE framework a natural choice for analyzing such correlated binary data. In this paper, we employ the GEE method to analyze combined unilateral and bilateral data and compare its performance with established likelihood-based procedures, including the likelihood ratio, Wald-type, and score tests. Specifically, we investigate the homogeneity of proportions for combined unilateral and bilateral data under both Rosner's and Donner's models. A recent study by Zhang and Ma~\cite{zhang2025analysis} compared the score test with the GEE method for equal proportion test in the combined data framework under Rosner's model. Our work extends this line of research by conducting a systematic comparison between GEE approach and the three likelihood-based methods under both Rosner's and Donner's models, providing a broader evaluation in this context.

The rest of the paper is organized as follows. Section~\ref{sec:methods} presents the MLE approaches using likelihood ratio, Wald-type and score statistics, as well as the GEE method for correlated binary outcomes and its implementation in \texttt{SAS}. Section~\ref{sec:results} reports the numerical results, including a simulation study (Section~\ref{subsec:simulation}) evaluating the empirical type I error rates and powers of these methods under different models, and two real world applications in otolaryngologic and ophthalmologic studies (Section~\ref{subsec:real-data}). Section~\ref{sec:conclusions} provides the discussion and conclusions.

\section{Methods}
\label{sec:methods}
Consider a study involving the combined unilateral and bilateral date, where in the $i$-th group $\brc{i=1,\ldots,g}$, there are $m_{+i}$ subjects who contribute data from both paired organs (bilateral), and $n_{+i}$ subjects who contribute data from one of the paired organs (unilateral), respectively. Let $m_{ri}$ be the number of bilateral subjects who have $r~\brc{r=0,1,2}$ organs cured or affected, and $n_{ri}$ be the number of unilateral subjects who have $r~\brc{r=0,1}$ organs cured or affected, such that 
\begin{align*}
  &m_{ri}=\sum_{j=1}^{m_{+i}}I\brc{Z_{ij1}+Z_{ij2}=r}, \quad r=0,1,2, \\
  &n_{ri}=\sum_{j=1}^{n_{+i}}I\brc{Z_{ijk}=r}, \quad r=0,1;~k=1\text{ or }2,
\end{align*}
where $Z_{ijk}$ denotes the response ($1$ cured or affected; $0$ otherwise) of the $k$-th paired organ ($k=1,2$) of the $j$-th subject in the $i$-th group. The data structure is summarized in Table~\ref{tab:data_struc},
where a subscript `$+$' indicates the summation over the corresponding index. 
\begin{table}[thpb]
    \centering
    \caption{Frequency table for number of cured or affected organs for subjects in $g$ groups.}
    \label{tab:data_struc}
    \begin{tabular}{cccccc}
    \toprule
     &\multicolumn{4}{c}{group} & \\
    \cline{2-5}
    \# of cured or affected organs &1 &2 &$\ldots$ &g &total \\
    \midrule
    0 &$m_{01}$ &$m_{02}$ &$\ldots$ &$m_{0g}$ &$m_{0+}$ \\
    1 &$m_{11}$ &$m_{12}$ &$\ldots$ &$m_{1g}$ &$m_{1+}$ \\
    2 &$m_{21}$ &$m_{22}$ &$\ldots$ &$m_{2g}$ &$m_{2+}$ \\
    total &$m_{+1}$ &$m_{+2}$ &$\ldots$ &$m_{+g}$ &$m_{++}$ \\
    \midrule
    0 &$n_{01}$ &$n_{02}$ &$\ldots$ &$n_{0g}$ &$n_{0+}$ \\
    1 &$n_{11}$ &$n_{12}$ &$\ldots$ &$n_{1g}$ &$n_{1+}$ \\
    total &$n_{+1}$ &$n_{+2}$ &$\ldots$ &$n_{+g}$ &$n_{++}$ \\
    \bottomrule
    \end{tabular}
\end{table}

The proportion of cured or affected organs in the $i$-th group is assumed to be $Pr\brc{Z_{ijk}=1}=\pi_i$. Given $m_{+i}$ and $n_{+i}$ in the $i$-th group, $\brc{m_{0i},m_{1i},m_{2i}}$ follows trinomial distribution and $\brc{n_{0i},n_{1i}}$ follows binomial distribution, i.e., 
\begin{equation}
  \brc{m_{0i},m_{1i},m_{2i}}\sim Trinomial\brc{m_{+i},p_{0i},p_{1i},p_{2i}}, \quad
  n_{1i} \sim Binomial\brc{n_{+i},\pi_i}, 
  \label{eq:distribution}
\end{equation}
where the joint probabilities $p_{ri}$'s ($r=0,1,2$) read
\begin{equation}
    \begin{aligned}
        &p_{2i}=Pr\brc{Z_{ij1}=1,Z_{ij2}=1}=E\brc{Z_{ij1}Z_{ij2}}=\pi_i\left[\pi_i+\brc{1-\pi_i}\corr\brc{Z_{ij1},Z_{ij2}}\right], \\
        &p_{1i}=\sum_{k=1}^2Pr\brc{Z_{ijk}=1,Z_{ij,3-k}=0}=2\brc{\pi_i-p_{2i}}=2\pi_i\brc{1-\pi_i}\left[1-\corr\brc{Z_{ij1},Z_{ij2}}\right], \\
        &p_{0i}=Pr\brc{Z_{ij1}=0,Z_{ij2}=0}=1-p_{1i}-p_{2i}=\brc{1-\pi_i}\left[1-\pi_i+\pi_i~\corr\brc{Z_{ij1},Z_{ij2}}\right],  
    \end{aligned}
    \label{eq:prob_ri}
\end{equation}
with $\corr\brc{Z_{ij1},Z_{ij2}}$ being the intra-subject correlation between the two responses from the $j$-th subject in the $i$-th group.

In what follows, we consider two parametric models proposed by Rosner~\cite{Rosner_1982} and Donner~\cite{Donner1989rhoModel} to address the intra-subject correlation for the binary data. Under Rosner's model, the conditional probability is specified as $Pr\brc{Z_{ijk}=1\mid Z_{ij,3-k}=1}=R\pi_i$, where $R$ is a scalar parameter. Based on this specification, the intra-subject correlation becomes $\corr\brc{Z_{ij1},Z_{ij2}}=\brc{R-1}\pi_i/\brc{1-\pi_i}$. Under Donner's model, a constant correlation $\rho$ is assumed across all $g$ groups, i.e., $\corr\brc{Z_{ij1},Z_{ij2}}=\rho$. Under either model, the joint probabilities in (\ref{eq:prob_ri}) can be written in terms of $\pi_i$ and the nuisance parameter $\kappa$ ($\kappa=R$ under Rosner's and $\kappa=\rho$ under Donner's model).

\subsection{MLE Approach}
\label{subsec:mle}
Let $\bbeta=\brc{\pi_1,\ldots,\pi_g,\kappa}^T$ be the vector of parameters. 
For given observation
$$\brc{m,n}=\brc{m_{01},m_{11},m_{2,1},\ldots,m_{0g},m_{1g},m_{2g},n_{01},n_{11},\ldots,n_{0g},n_{1g}},$$
the log-likelihood function reads 
\begin{equation}
%  l\brc{\pi_1,\ldots,\pi_g,\kappa|\brc{m,n}}
 l\brc{\bbeta} =\sum_{i=1}^g\sum_{r=0}^2m_{ri}\log\brc{p_{ri}}+\sum_{i=1}^g\left[n_{0i}\log\brc{1-\pi_i}+n_{1i}\log\brc{\pi_i}\right]+\text{const},
    \label{eq:ll:generic}
\end{equation}
where the term `const" denotes a constant depending on $\brc{m,n}$.

Our interest is the homogeneity test of proportions across the $g$ groups. Thus, the hypotheses are
\begin{equation}
H_0:~\pi_1=\cdots=\pi_g=\pi, 
\quad
\text{ versus } 
\quad
H_1:~\text{some of $\pi_i$'s are not equal.}
\label{eq:hypotheses}
\end{equation}

The MLEs $\hat{\bbeta}_0$ under $H_0$ can be solved analytically, while the MLEs $\hat{\bbeta}_1$ under $H_1$ have no closed-form solutions and must be computed using an iterative method. Detailed procedures for obtaining $\hat{\bbeta}_0$ and $\hat{\bbeta}_1$ can be found in the works of Ma and Wang K.~\cite{ma2021testing} and Ma and Wang H.~\cite{ma2022testing}, corresponding to Rosner's and Donner's models, respectively. Based on these MLEs, three likelihood-based test statistics are considered and defined as follows.

\paragraph{Likelihood Ratio (LR) Test}
\begin{equation}
  Q_{LR}=2\left[l\brc{\hat{\bbeta}_1}-l\brc{\hat{\bbeta}_0}\right].
  \label{eq:LR}
\end{equation}

\paragraph{Wald-type Test}
\begin{equation}
  Q_W=\left.\brc{\bbeta^T\bC}\left[\bC^T\bI\brc{\bbeta}\bC\right]^{-1}\brc{\bC^T\bbeta}\right|_{\bbeta=\hat{\bbeta}_0},
  \label{eq:wald}
\end{equation}
where $\bC^T$ is a $\brc{g-1}\times\brc{g+1}$ hypothesis matrix with $\brc{\bC^T}_{ii}=1$, $\brc{\bC^T}_{i,i+1}=-1$ for $i=1,\ldots,g-1$, and $\brc{\bC^T}_{ij}=0$ otherwise; $I\brc{\bbeta}$ is the $\brc{g+1}\times\brc{g+1}$ Fisher's information matrix, and its explicit forms under Rosner's and Donner's models are given in Appendices in the works of Ma and Wang K.~\cite{ma2021testing} and Ma and Wang H.~\cite{ma2022testing}, respectively.
%% $$
%% \boldsymbol{C}^T=\left[
%% \begin{array}{rrrrrr}
%%     1 &-1 &0 &0 &\cdots &0 \\
%%     0 &1 &-1 &0 &\cdots &0 \\
%%     \vdots &\vdots &\ddots &\ddots &\cdots &\vdots \\
%%     0 &0 &0 &1 &-1 &0    
%% \end{array}
%% \right]_{\brc{g-1}\times\brc{g+1}}, 
%% $$
%% such that the null hypothesis in matrix form reads $H_0:~\boldsymbol{C}^T\bbeta=\boldsymbol{0}$.

\paragraph{Score Test}
\begin{equation}
  Q_S=\left.\bU\bI^{-1}\brc{\bbeta}\bU^T\right|_{\bbeta=\hat{\bbeta}},
  \label{eq:score}
\end{equation}
where $\bU=\brc{\partial l/\partial\pi_1,\ldots,\partial l/\partial\pi_g,0}$ is the score function, and $\bI^{-1}\brc{\bbeta}$ is the inverse of the Fisher's information matrix $\bI\brc{\bbeta}$.

It can be shown that the above three test statistics asymptotically follow the chi-square distribution with degrees of freedom $df=g-1$ under $H_0$, i.e., $Q_{LR},Q_W,Q_S\stackrel{d}{\to}\chi^2_{g-1}$.

\subsection{GEE Approach}
\label{subsec:gee}
Treating the bilateral observations as repeated measurements on the same subject, the GEE method can be employed to perform equal proportion test with the mean model $\mathbb{E}\brc{\bZ_{ij}}=\bmu_{ij}=\pi_i\mathbbm{1}_{k_{ij}}$~\footnote{$\mathbbm{1}_{k_{ij}}$ denotes a $k_{ij}\times1$ column vector of ones.}, and the estimating equations $\sum_{j=1}^{n_i}\partial\bmu_{ij}^T/\partial\pi_i\bV_{ij}^{-1}\brc{\bZ_{ij}-\bmu_{ij}}=0$, where $\bZ_{ij}$ denotes the response(s) of the $j$-th subject in the $i$-th group, which is a $2\times1$ vector for bilateral data and a scalar for unilateral data, $k_{ij}=2$ for bilateral data and $k_{ij}=1$ for unilateral data, and $\bV_{ij}=\bA_{ij}^{1/2}\bR_{ij}\bA_{ij}^{1/2}$ is the ``working covariance'' matrix. Here, $\bR_{ij}$ is the ``working correlation'' and $\bA_{ij}$ is the diagonal matrix of marginal variances of $\bZ_{ij}$.
The working correlation matrix $\boldsymbol{R}_{ij}$ generally depends on unknown parameters that must be estimated. In the case of paired organ data, $R_{ij}$ contains only a single parameter. Various working correlation structures can be specified, such as independent, exchangeable, and unstructured correlations. Estimates for the parameters of both the mean model and the working covariance are obtained through an iterative procedure based on the estimating equations (see, e.g., the \texttt{GENMOD} procedure in the \texttt{SAS} user's guide~\cite{sasgenmod}). 

Several statistical software systems can implement the GEE approach for analyzing combined unilateral and bilateral data. In this study, we use the \texttt{SAS GENMOD} procedure. Table \ref{tab:stack:data} illustrates the structure of the combined unilateral and bilateral data for the case $g=2$ in a single simulation. 
\begin{table}[thpb]
    \centering
    \caption{The structure of stacked data ($g=2$) in the first simulation (\texttt{replicate=1}).}
    \label{tab:stack:data}
    \begin{tabular}{ccccc}
    \toprule
        sub\_id &response &group &count &replicate \\
        \midrule
        1 &0 &1 &$m_{01}$ &1 \\
        1 &0 &1 &$m_{01}$ &1 \\
        2 &1 &1 &$m_{11}$ &1 \\
        2 &0 &1 &$m_{11}$ &1 \\
        3 &1 &1 &$m_{21}$ &1 \\
        3 &1 &1 &$m_{21}$ &1 \\
        4 &0 &2 &$m_{02}$ &1 \\
        4 &0 &2 &$m_{02}$ &1 \\
        5 &1 &2 &$m_{12}$ &1 \\
        5 &0 &2 &$m_{12}$ &1 \\
        6 &1 &2 &$m_{22}$ &1 \\
        6 &1 &2 &$m_{22}$ &1 \\
        \midrule
        7 &0 &1 &$n_{01}$ &1 \\
        8 &1 &1 &$n_{11}$ &1 \\
        9 &0 &2 &$n_{02}$ &1 \\
        10 &1 &2 &$n_{12}$ &1 \\
        \bottomrule
    \end{tabular}
\end{table}

An example of \texttt{SAS} code invoking the \texttt{GENMOD} procedure for $g=2$ is presented in Table~\ref{tab:genmod}. In this example, the link function is the identity link specified with \texttt{link=identity}, and the unstructured working correlation matrix is specified with \texttt{type=un}. In addition, the \texttt{by replicate} statement repeats the analysis for each simulation case, and the \texttt{contrast} statememnt enables pairwise evaluation of differences, corresponding to the equal proportion test. Note that with \texttt{repeated} statement, the default statistic provided by the \texttt{contrast} statement is a score test $Q_{GS}$ computed based on the generalized score function~\cite{sasgenmod}. 
\begin{table}[thpb]
  \centering
  \caption{Pseudo \texttt{SAS} code using \texttt{PROC GENMOD}}
  \label{tab:genmod}
%% \begin{tcolorbox}[text width=0.71\textwidth]
%% \begin{minted}{sas}
%% * stacked data (g=2) file w/ name <data_g2_stacked>;
%% * var names same with those in the above table (Table 2);
%% proc genmod data=data_g2_stacked descending;
%% freq count;
%% class group sub_id;
%% model response=group / link=identity dist=bin;
%% repeated subject=sub_id / type=un corrw;
%% by replicate;
%% contrast 'group' group 1 -1;
%% run;
%% \end{minted}        
%% \end{tcolorbox}
\vspace{3pt}
\includegraphics[scale=0.57]{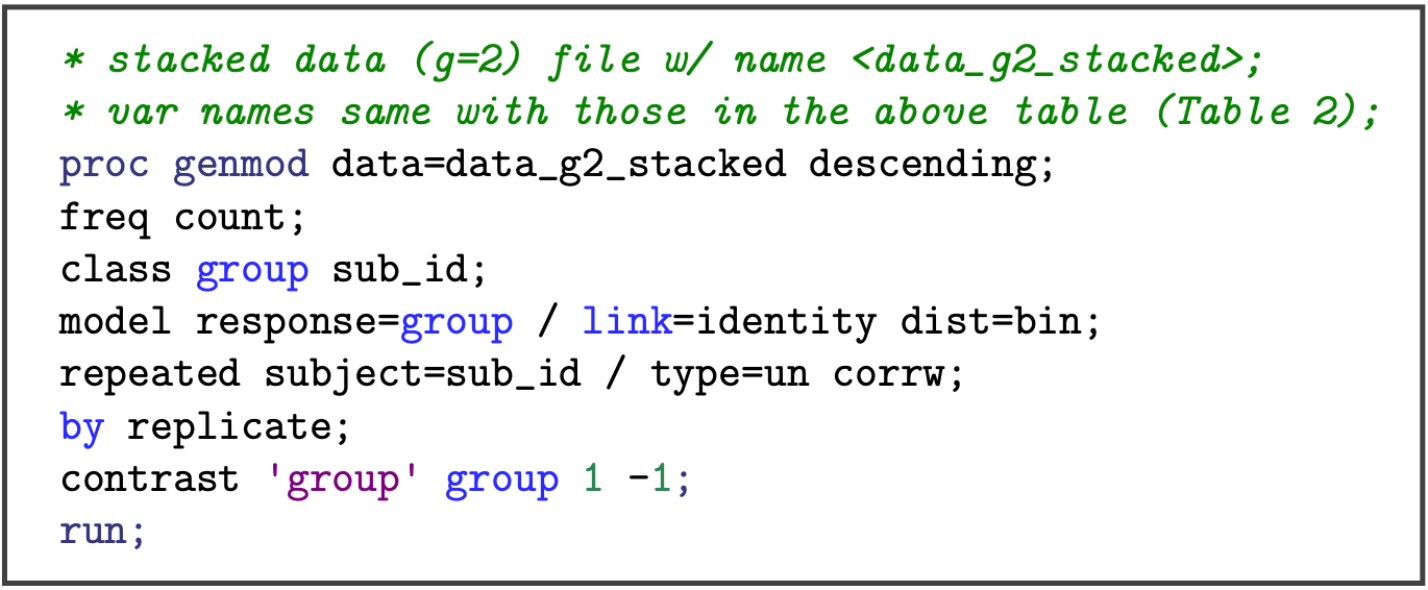}
\end{table}

\section{Results}
\label{sec:results}

\subsection{Simulation Study}
\label{subsec:simulation}
We conduct a simulation study to assess the performance of the three likelihood-based test statistics along with the GEE-based generalized score test, by investigating the empirical type I error and powers respectively under Rosner's and Donner's model. 
We consider equal and unequal sample size for $g=2,4,8$. Specifically, for equal sample size design, we set
$$
m_{+1}=\cdots=m_{+g}=n_{+1}=\cdots=n_{+g}=20,40,
$$ 
and for unequal sample size design, we set
$$
\brc{m_{+1},\ldots,m_{+g}}=\brc{n_{+1},\ldots,n_{+g}}=\brc{20,40},~\brc{20,20,40,40},~\brc{20,20,30,30,40,40,50,50}.
$$
%% we set the sample size for $m_{+i}$ and $n_{+i}$ as shown in the table below. 
%% \begin{table}[tbhp]
%%     \centering
%%     \caption{Design of Unequal Sample Size}
%%     \begin{tabular}{c|c}
%%     \hline
%%         Design &$\brc{m_{+1},m_{+2},\ldots,m_{+g}}
%% =\brc{n_{+1},n_{+2},\ldots,n_{+g}}$  \\
%%         \hline
%%         $U_2$ &$\brc{20,40}$ \\
%%         $U_4$ &$\brc{20,20,40,40}$ \\
%%         $U_8$ &$\brc{20,20,30,30,40,40,50,50}$ \\
%%     \end{tabular}
%%     \label{tab:sample-size}
%% \end{table}

%Note that for simplicity, the designs of sample size $\brc{m_{+1},\ldots,m_{+g}}$ and $\brc{n_{+1},\ldots,n_{+g}}$ are denoted by $\brc{m,n}$ in the following tables.

\subsubsection{Empirical Type I Error}
To compute the empirical type I error rates, we generate the dataset according to the distributions in (\ref{eq:distribution}) with the above sample size designs under $H_0:~\pi_1=\cdots=\pi_g=\pi_0$ with a pre-specified $\pi_0=0.3,0.5$. Additionally, we set the intra-subject correlation to be $\rho_0=0.4,0.5,0.7$. Thus, in data generation, we set $R=R_0=\brc{1-\pi_0}\rho_0/\pi_0+1$ under Rosner's model, and $\rho=\rho_0$ under Donner's model. 

After generating the dataset with each sample size design and parameter configuration, the three likelihood-based test statistics $Q_{LR},Q_W,Q_S$ in (\ref{eq:LR}) - (\ref{eq:score}), and the GEE-based test statistic denoted by $Q_{GS}$ are computed accordingly. The null hypothesis $H_0:~\pi_1=\cdots=\pi_g$ is rejected if $\text{Test Statistic}>\chi_{1-\alpha,~g-1}^2$, where $\chi_{1-\alpha,~g-1}^2$ is the quantile function of $\chi_{g-1}^2$ valued at $1-\alpha$. The above simulation is replicated for $N=50000$ times. Then the empirical type I error rate is calculated as 
\begin{equation}
    \widehat{\mathrm{TIE}}=\frac{\sum_{k=1}^NI\brc{Q_i^{H_0}>\chi_{1-\alpha,~g-1}^2}}{N}, 
\end{equation}
where $Q_i^{H_0}$ denotes the test statistic resulting from the dataset generated under $H_0$, and the subscript ``$i$" represents the type of tests for $i=LR,W,S,GS$.

The results for the empirical type I error with each sample size design and parameter configuration are presented in Table~\ref{tab:TIE:rosner} under Rosner's model, and in Table~\ref{tab:TIE:donner} under Donner's model, respectively, where the liberal results are highlighted in boldface~\footnote{The results are classified as \textit{liberal} if $\widehat{\mathrm{TIE}}>0.06$, as \textit{robust} if $0.04\le\widehat{\mathrm{TIE}}\le0.06$, and as \textit{conservative} if $\widehat{\mathrm{TIE}}<0.04$.}.
For simplicity, the designs of sample size $\brc{m_{+1},\ldots,m_{+g}}$ and $\brc{n_{+1},\ldots,n_{+g}}$ are referred to as $\brc{m,n}$ in Tables~\ref{tab:TIE:rosner} and \ref{tab:TIE:donner}. The two equal sample size designs, where $m_{+1}=\cdots=m_{+g}=n_{+1}=\cdots=n_{+g}=20,40$, are denoted as $E_1$ and $E_2$, respectively. The unequal sample size designs are labeled as $U_g$ for $g=2,4,8$. For instance, $U_2$ refers to the design where $\brc{m_{+1},m_{+2}}=\brc{n_{+1},n_{+2}}=\brc{20,40}$. 
Note that these notations also apply to the tables presented for power calculation.
%TIE under Rosner's model
{\scriptsize
\begin{longtblr}[
   caption={The empirical type I error rates (in \%) under Rosner's model for different test procedures, under $H_0:~\pi_1=\cdots=\pi_g=\pi_0$ at the nominal level of $\alpha=0.05$.},
   label={tab:TIE:rosner}
 ]{
   cell{1}{4}={c=4}{halign=c},
   cell{1}{8}={c=4}{halign=c},
   cell{1}{12}={c=4}{halign=c}
 }
 \toprule
 $\brc{m,n}$ &$\pi_0$ &$\rho$ &$g=2$ & & & &$g=4$ & & & &$g=8$ & & & \\
 \cline{4-15}
  & & &$Q_{LR}$ &$Q_W$ &$Q_S$ &$Q_{GS}$ &$Q_{LR}$ &$Q_W$ &$Q_S$ &$Q_{GS}$ &$Q_{LR}$ &$Q_W$ &$Q_S$ &$Q_{GS}$ \\
 \midrule
 $E_1$     &0.3      &0.4      &{\bf  6.05}   &{\bf  6.04}   & 4.98         & 5.21         & 5.81         &{\bf  7.58}   & 4.50         & 5.18         & 5.82         &{\bf 10.35}   & 4.33         & 5.33          \\ 
           &         &0.5      &{\bf  6.30}   & 5.72         & 4.92         & 5.20         &{\bf  6.16}   &{\bf  7.65}   & 4.59         & 5.13         &{\bf  6.20}   &{\bf 11.44}   & 4.46         & 5.42          \\ 
           &         &0.7      & 5.81         & 4.12         & 5.38         & 5.13         & 5.32         & 5.47         & 4.13         & 5.22         & 5.16         &{\bf  8.03}   & 3.91         & 5.53          \\ 
           &0.5      &0.4      & 5.98         &{\bf  6.06}   & 5.04         & 4.99         &{\bf  6.09}   &{\bf  7.94}   & 4.94         & 4.95         &{\bf  6.30}   &{\bf 10.62}   & 4.92         & 4.83          \\ 
           &         &0.5      &{\bf  6.09}   & 5.68         & 5.12         & 5.12         &{\bf  6.14}   &{\bf  7.69}   & 4.88         & 4.96         &{\bf  6.40}   &{\bf 11.24}   & 4.81         & 4.98          \\ 
           &         &0.7      & 5.27         & 3.66         & 5.52         & 5.07         & 4.92         & 4.80         & 4.31         & 5.12         & 4.78         &{\bf  6.82}   & 4.03         & 4.92          \\ 
 $E_2$     &0.3      &0.4      & 5.63         & 5.62         & 5.03         & 5.00         & 5.88         &{\bf  6.91}   & 5.04         & 5.24         & 5.88         &{\bf  8.21}   & 4.96         & 5.22          \\ 
           &         &0.5      & 5.82         & 5.56         & 5.04         & 5.15         & 5.95         &{\bf  6.93}   & 4.98         & 5.17         &{\bf  6.18}   &{\bf  9.14}   & 4.94         & 5.36          \\ 
           &         &0.7      & 5.97         & 4.47         & 5.07         & 5.04         &{\bf  6.08}   &{\bf  6.66}   & 4.81         & 5.26         &{\bf  6.35}   &{\bf 10.87}   & 4.83         & 5.04          \\ 
           &0.5      &0.4      & 5.50         & 5.59         & 5.06         & 5.10         & 5.62         &{\bf  6.46}   & 5.09         & 5.08         & 5.68         &{\bf  7.43}   & 5.04         & 5.04          \\ 
           &         &0.5      & 5.50         & 5.29         & 5.04         & 4.99         & 5.63         &{\bf  6.42}   & 5.02         & 5.03         & 5.62         &{\bf  7.71}   & 4.79         & 4.75          \\ 
           &         &0.7      & 5.82         & 4.67         & 5.12         & 5.13         & 5.93         &{\bf  6.73}   & 4.92         & 5.13         &{\bf  6.02}   &{\bf 10.26}   & 4.84         & 4.95          \\ 
 $U_g$     &0.3      &0.4      & 5.67         &{\bf  6.12}   & 4.79         & 5.08         & 5.67         &{\bf  7.63}   & 4.69         & 5.23         & 5.91         &{\bf  9.53}   & 4.84         & 5.41          \\ 
           &         &0.5      & 6.00         &{\bf  6.28}   & 4.95         & 5.28         & 5.99         &{\bf  8.03}   & 4.71         & 5.09         &{\bf  6.23}   &{\bf 10.84}   & 4.83         & 5.52          \\ 
           &         &0.7      & 5.61         & 4.71         & 5.05         & 5.18         & 5.48         &{\bf  6.19}   & 4.50         & 5.32         & 5.94         &{\bf 10.33}   & 4.56         & 5.43          \\ 
           &0.5      &0.4      & 5.80         &{\bf  6.28}   & 5.18         & 5.18         & 5.82         &{\bf  7.58}   & 5.01         & 5.01         & 5.85         &{\bf  8.83}   & 4.96         & 4.94          \\ 
           &         &0.5      & 5.75         &{\bf  6.15}   & 5.04         & 5.21         & 5.87         &{\bf  7.76}   & 4.96         & 4.94         & 5.87         &{\bf  9.75}   & 4.86         & 4.84          \\ 
           &         &0.7      & 5.25         & 4.77         & 4.67         & 5.03         & 5.31         &{\bf  6.46}   & 4.61         & 4.96         & 5.52         &{\bf 10.38}   & 4.51         & 4.91          \\ 
 \bottomrule
 \end{longtblr}
}

%TIE under Donner's model
{\scriptsize
 \begin{longtblr}[
   caption={The empirical type I error rates (in \%) under Donner's model for different test procedures, under $H_0:~\pi_1=\cdots=\pi_g=\pi_0$ at the nominal level $\alpha=0.05$.},
   label={tab:TIE:donner}
 ]{
   cell{1}{4}={c=4}{halign=c},
   cell{1}{8}={c=4}{halign=c},
   cell{1}{12}={c=4}{halign=c}
 }
 \toprule
 $\brc{m,n}$ &$\pi_0$ &$\rho$ &$g=2$ & & & &$g=4$ & & & &$g=8$ & & & \\
 \cline{4-15}
  & & &$Q_{LR}$ &$Q_W$ &$Q_S$ &$Q_{GS}$ &$Q_{LR}$ &$Q_W$ &$Q_S$ &$Q_{GS}$ &$Q_{LR}$ &$Q_W$ &$Q_S$ &$Q_{GS}$ \\
 \midrule
 $E_1$     &0.3      &0.4      & 5.28         & 5.67         & 5.08         & 5.16         & 5.05         &{\bf  6.42}   & 4.67         & 5.11         & 5.18         &{\bf  7.74}   & 4.64         & 5.44          \\ 
           &         &0.5      & 5.27         & 5.64         & 5.06         & 5.18         & 5.27         &{\bf  6.67}   & 4.87         & 5.24         & 5.10         &{\bf  7.80}   & 4.59         & 5.27          \\ 
           &         &0.7      & 5.27         & 5.66         & 5.05         & 5.17         & 5.21         &{\bf  6.61}   & 4.81         & 5.25         & 5.24         &{\bf  8.02}   & 4.71         & 5.47          \\ 
           &0.5      &0.4      & 5.23         & 5.74         & 5.04         & 5.03         & 5.24         &{\bf  6.58}   & 4.93         & 4.88         & 5.47         &{\bf  7.71}   & 4.98         & 4.92          \\ 
           &         &0.5      & 5.37         & 5.90         & 5.22         & 5.21         & 5.33         &{\bf  6.63}   & 4.98         & 4.98         & 5.29         &{\bf  7.58}   & 4.80         & 4.78          \\ 
           &         &0.7      & 5.21         & 5.67         & 5.08         & 5.08         & 5.25         &{\bf  6.62}   & 4.94         & 4.95         & 5.48         &{\bf  7.91}   & 5.00         & 5.00          \\ 
 $E_2$     &0.3      &0.4      & 5.18         & 5.45         & 5.07         & 5.07         & 5.11         & 5.86         & 4.90         & 5.04         & 5.24         &{\bf  6.50}   & 4.98         & 5.22          \\ 
           &         &0.5      & 5.21         & 5.42         & 5.09         & 5.12         & 5.18         & 5.90         & 4.96         & 5.12         & 5.21         &{\bf  6.54}   & 4.93         & 5.14          \\ 
           &         &0.7      & 5.17         & 5.35         & 5.05         & 5.03         & 5.21         & 5.90         & 5.03         & 5.08         & 5.28         &{\bf  6.58}   & 4.96         & 5.21          \\ 
           &0.5      &0.4      & 5.09         & 5.33         & 5.00         & 5.00         & 5.19         & 5.85         & 5.01         & 5.03         & 5.25         &{\bf  6.30}   & 5.01         & 5.00          \\ 
           &         &0.5      & 5.10         & 5.39         & 5.04         & 5.06         & 5.08         & 5.71         & 4.94         & 4.95         & 5.27         &{\bf  6.33}   & 5.01         & 5.00          \\ 
           &         &0.7      & 5.22         & 5.46         & 5.16         & 5.16         & 5.10         & 5.77         & 4.96         & 4.98         & 5.16         &{\bf  6.34}   & 4.92         & 4.90          \\ 
 $U_g$     &0.3      &0.4      & 5.23         & 5.64         & 5.05         & 5.18         & 5.12         &{\bf  6.30}   & 4.81         & 5.21         & 5.21         &{\bf  6.92}   & 4.85         & 5.29          \\ 
           &         &0.5      & 5.08         & 5.56         & 4.94         & 5.12         & 5.18         &{\bf  6.40}   & 4.88         & 5.32         & 5.21         &{\bf  7.03}   & 4.86         & 5.37          \\ 
           &         &0.7      & 5.06         & 5.52         & 4.88         & 5.07         & 5.04         &{\bf  6.30}   & 4.79         & 5.18         & 5.22         &{\bf  7.13}   & 4.83         & 5.41          \\ 
           &0.5      &0.4      & 5.17         & 5.61         & 5.02         & 5.03         & 5.31         &{\bf  6.28}   & 5.04         & 5.00         & 5.20         &{\bf  6.55}   & 4.89         & 4.86          \\ 
           &         &0.5      & 5.21         & 5.67         & 5.07         & 5.10         & 5.24         &{\bf  6.33}   & 5.00         & 5.00         & 5.26         &{\bf  6.75}   & 4.96         & 4.95          \\ 
           &         &0.7      & 5.24         & 5.76         & 5.13         & 5.12         & 5.18         &{\bf  6.30}   & 4.95         & 4.97         & 5.29         &{\bf  6.79}   & 5.00         & 5.00          \\ 
 \bottomrule
 \end{longtblr}
}

Under Rosner's model, as shown in Table \ref{tab:TIE:rosner}, when $g=2$, LR tests produce a few liberal results for the sample size design $E_1$, while Wald tests produce more liberal results for the sample size design $E_2$ and $U_2$. As $g=4$, the number of liberal results from Wald tests grows dramatically for all sample size designs ($E_1$, $E_2$ and $U_4$), in contrast to LR tests, which show roughly the same number of the liberal, with one liberal result for sample design $E_2$. When $g=8$, every empirical type I error rate from Wald tests is liberal, with over half of them being considerably large (rate $>0.10$) for all sample size designs. In other words, as the number of groups increases, LR tests produce moderately more liberal results, while Wald tests show a dramatic increase in both the number and amplitude of the liberal results. Additionally, there is no apparent association between the empirical type I error rates and the null proportion $\pi_0$ or the intra-subject correlation $\rho$ for either LR or Wald tests. Unlike LR or Wald tests, score tests produce robust empirical type I error rates regardless of the simulation designs. Moreover, the results from score tests are consistent with those from GEE method and are closer to the nominal level, as compared to LR and Wald tests. 

Under Donner's model, as shown in Table \ref{tab:TIE:donner}, Wald tests begin to produce liberal results when $g=4$ for sample size designs $E_1$ and $U_4$. As $g=8$, all results from Wald tests are liberal for all designs. However, compared to the results under Rosner's model, the liberal results under Donner's model have smaller amplitudes (rate $\lesssim0.08$). All other tests produce robust empirical type I error rates regardless of the simulation designs, with score tests and GEE tests yielding similar results that are closer to nominal level results, as compared to LR tests.

\subsubsection{Powers}
The powers are calculated in a similar way as it is done for the empirical type I error. The difference is that instead of generating the dataset under $H_0$, we generate them under certain alternative hypothesis. There are two alternatives $H_{1A}$ and $H_{1B}$ used for power calculation, which are shown below.
\begin{align*}
  &H_{1A}:~\brc{\pi_1,\ldots,\pi_g}=\brc{0.25,0.4},~\brc{0.25,0.3,0.35,0.4},~\brc{0.25,0.3,0.35,0.4,0.25,0.3,0.35,0.4}, \\
  &H_{1B}:~\brc{\pi_1,\ldots,\pi_g}=\brc{0.2,0.4},~\brc{0.2,0.2,0.4,0.4},~\brc{0.2,0.2,0.4,0.4,0.2,0.2,0.4,0.4}, 
\end{align*}
for $g=2,4,8$.
%% which are shown in the table below. 
%% \begin{table}[tbhp]
%%     \centering
%%     \caption{Two Alternative Hypotheses $H_{1A}$ and $H_{1B}$ for $g=2,4,8$.}
%%     \begin{tabular}{c|cc}
%%     \hline
%%          &$H_{1A}$ &$H_{1B}$ \\
%%          \hline
%%         g &$\brc{\pi_1,\ldots,\pi_g}$ &$\brc{\pi_1,\ldots,\pi_g}$ \\
%%         \hline
%%         2 &$\brc{0.25,0.4}$ &$\brc{0.2,0.4}$ \\
%% %        3 &$\brc{0.25,0.3,0.4}$ &$\brc{0.2,0.3,0.4}$ \\
%%         4 &$\brc{0.25,0.3,0.35,0.4}$ &$\brc{0.2,0.2,0.4,0.4}$ \\
%% %        5 &$\brc{0.25,0.3,0.3,0.35,0.4}$ &$\brc{0.2,0.2,0.3,0.4,0.4}$ \\
%% %        6 &$\brc{0.25,0.3,0.3,0.35,0.35,0.4}$ &$\brc{0.2,0.2,0.3,0.3,0.4,0.4}$ \\
%% %        7 &$\brc{0.25,0.25,0.3,0.3,0.35,0.35,0.4}$ &$\brc{0.2,0.2,0.3,0.3,0.3,0.4,0.4}$ \\
%%         8 &$\brc{0.25,0.3,0.35,0.4,0.25,0.3,0.35,0.4}$ &$\brc{0.2,0.2,0.4,0.4,0.2,0.2,0.4,0.4}$
%%     \end{tabular}
%%     \label{tab:H1}
%% \end{table}
%% The test statistics based on the MLEs and GEE method are then computed accordingly in each data simulation. The power can be calculated after $N=50000$ times of simulations as 
%% \begin{equation}
%%     \widehat{\mathrm{Power}}=\frac{\sum_{k=1}^NI\brc{Q_i^{H_1}>\chi_{1-\alpha,~g-1}^2}}{N}, 
%% \end{equation}
%% where $Q_i^{H_1}$ denotes the test statistic resulting from dataset generated under the alternative (either $H_{1A}$ or $H_{1B}$) for $i=LR,W,S,X^2$. 
Tables~\ref{tab:power:rosner} and \ref{tab:power:donner} present the results for powers under under $H_{1A}$ and $H_{1B}$ within Rosner's and Donner's model, respectively.

{\scriptsize
 \begin{longtblr}[
   caption={The powers (in $\%$) under Rosner's model at the nominal level of $\alpha=0.05$.},
   label={tab:power:rosner}
 ]{
   cell{1}{3}={c=4}{halign=c},
   cell{1}{7}={c=4}{halign=c},
   cell{1}{11}={c=4}{halign=c},
   cell{3}{1}={c=14}{halign=c},
   cell{13}{1}={c=14}{halign=c}
 }
 \toprule
 $\brc{m,n}$ &$R$ &$g=2$ & & & &$g=4$ & & & &$g=8$ & & & \\
  \cline{3-14}
  & &$Q_{LR}$ &$Q_W$ &$Q_S$ &$Q_{GS}$ &$Q_{LR}$ &$Q_W$ &$Q_S$ &$Q_{GS}$ &$Q_{LR}$ &$Q_W$ &$Q_S$ &$Q_{GS}$ \\
  \midrule
  under $H_{1A}$ & & & & & & & & & & & & & \\
  \midrule
 $E_1$     &1.4      & 39.64        & 41.17        & 36.75        & 38.09        & 28.66        & 32.19        & 25.97        & 27.02        & 39.13        & 45.71        & 35.80        & 37.40         \\ 
           &1.5      & 39.73        & 41.04        & 36.53        & 37.10        & 29.18        & 32.61        & 25.94        & 26.58        & 39.38        & 46.04        & 35.85        & 36.37         \\ 
           &1.7      & 41.08        & 41.32        & 37.23        & 35.59        & 30.31        & 33.68        & 26.46        & 25.30        & 41.91        & 49.13        & 37.24        & 34.99         \\ 
 $E_2$     &1.4      & 66.75        & 67.60        & 65.07        & 64.85        & 54.96        & 57.15        & 53.18        & 52.38        & 74.61        & 77.30        & 72.97        & 71.72         \\ 
           &1.5      & 66.78        & 67.64        & 64.94        & 63.66        & 55.15        & 57.28        & 53.17        & 51.27        & 74.57        & 77.13        & 72.69        & 70.36         \\ 
           &1.7      & 68.16        & 68.49        & 66.07        & 61.75        & 56.59        & 58.46        & 53.89        & 48.70        & 76.30        & 78.80        & 74.05        & 67.80         \\ 
 $U_g$     &1.4      & 48.71        & 52.14        & 43.92        & 49.56        & 37.29        & 42.89        & 32.59        & 38.36        & 65.87        & 70.42        & 62.47        & 64.40         \\ 
           &1.5      & 48.71        & 51.98        & 43.23        & 48.58        & 37.54        & 43.09        & 32.33        & 37.56        & 65.53        & 70.27        & 61.88        & 62.71         \\ 
           &1.7      & 50.51        & 52.65        & 43.85        & 47.36        & 39.62        & 44.39        & 33.53        & 36.48        & 67.77        & 72.42        & 63.31        & 60.47         \\ 
  \midrule
  under $H_{1B}$ & & & & & & & & & & & & & \\
  \midrule
 $E_1$      &1.4      & 62.89        & 64.38        & 59.41        & 62.37        & 77.73        & 80.12        & 74.67        & 77.27        & 93.59        & 94.97        & 92.31        & 93.45         \\ 
            &1.5      & 62.25        & 63.63        & 58.09        & 60.69        & 77.70        & 79.95        & 74.23        & 76.05        & 93.53        & 94.93        & 92.05        & 92.79         \\ 
            &1.7      & 63.10        & 63.90        & 57.98        & 58.73        & 78.34        & 80.13        & 73.87        & 73.49        & 94.23        & 95.41        & 92.17        & 91.33         \\ 
 $E_2$      &1.4      & 90.45        & 90.94        & 89.53        & 89.83        & 98.31        & 98.44        & 98.07        & 97.90        & 99.96        & 99.97        & 99.96        & 99.95         \\ 
            &1.5      & 90.51        & 91.05        & 89.36        & 89.38        & 98.24        & 98.41        & 97.90        & 97.66        & 99.96        & 99.96        & 99.96        & 99.95         \\ 
            &1.7      & 90.63        & 90.97        & 89.13        & 87.70        & 98.26        & 98.37        & 97.85        & 96.96        & 99.96        & 99.97        & 99.95        & 99.89         \\ 
 $U_g$      &1.4      & 75.23        & 78.40        & 70.28        & 76.91        & 89.25        & 91.92        & 85.35        & 90.52        & 99.81        & 99.88        & 99.74        & 99.82         \\ 
            &1.5      & 75.05        & 78.09        & 68.80        & 75.88        & 89.24        & 91.88        & 84.53        & 90.11        & 99.82        & 99.88        & 99.72        & 99.79         \\ 
            &1.7      & 75.56        & 77.93        & 67.69        & 74.06        & 89.76        & 92.02        & 83.76        & 88.49        & 99.84        & 99.88        & 99.71        & 99.69         \\ 
\bottomrule  
 \end{longtblr}
}

{\scriptsize
 \begin{longtblr}[
   caption={The powers (in $\%$) under Donner's model at the nominal level of $\alpha=0.05$.},
   label={tab:power:donner}
 ]{
   cell{1}{3}={c=4}{halign=c},
   cell{1}{7}={c=4}{halign=c},
   cell{1}{11}={c=4}{halign=c},
   cell{3}{1}={c=14}{halign=c},
   cell{13}{1}={c=14}{halign=c}
 }
 \toprule
 $\brc{m,n}$ &$\rho$ &$g=2$ & & & &$g=4$ & & & &$g=8$ & & & \\
 \cline{3-14}
  & &$Q_{LR}$ &$Q_W$ &$Q_S$ &$Q_{GS}$ &$Q_{LR}$ &$Q_W$ &$Q_S$ &$Q_{GS}$ &$Q_{LR}$ &$Q_W$ &$Q_S$ &$Q_{GS}$ \\
 \midrule
 under $H_{1A}$ & & & & & & & & & & & & & \\
 \midrule
 $E_1$     &0.4      & 35.94        & 37.18        & 35.40        & 35.23        & 25.81        & 29.17        & 24.71        & 25.23        & 35.33        & 41.62        & 33.67        & 34.23         \\ 
           &0.5      & 34.84        & 36.06        & 34.26        & 34.16        & 25.39        & 28.61        & 24.47        & 24.73        & 34.59        & 40.93        & 32.88        & 33.69         \\ 
           &0.7      & 33.03        & 34.23        & 32.53        & 32.48        & 23.34        & 26.72        & 22.35        & 22.81        & 32.01        & 38.44        & 30.30        & 31.07         \\ 
 $E_2$     &0.4      & 61.19        & 61.86        & 60.90        & 60.67        & 49.51        & 51.27        & 48.80        & 48.51        & 68.04        & 70.78        & 67.18        & 66.98         \\ 
           &0.5      & 59.94        & 60.62        & 59.63        & 59.27        & 47.34        & 49.40        & 46.65        & 46.46        & 66.07        & 68.87        & 65.15        & 65.08         \\ 
           &0.7      & 57.31        & 57.94        & 57.01        & 56.69        & 44.86        & 46.94        & 44.27        & 44.23        & 62.69        & 65.80        & 61.81        & 61.72         \\ 
 $U_g$     &0.4      & 45.17        & 48.37        & 43.74        & 46.46        & 34.55        & 39.53        & 32.68        & 35.99        & 59.25        & 64.06        & 57.70        & 59.50         \\ 
           &0.5      & 43.72        & 47.14        & 42.33        & 45.18        & 33.06        & 38.47        & 31.10        & 34.54        & 57.87        & 63.02        & 56.18        & 58.05         \\ 
           &0.7      & 41.54        & 44.73        & 39.99        & 42.86        & 30.78        & 36.29        & 28.73        & 32.59        & 53.95        & 59.31        & 52.18        & 54.52         \\ 
 \midrule
 under $H_{1B}$ & & & & & & & & & & & & & \\
 \midrule
 $E_1$     &0.4      & 59.38        & 60.67        & 58.72        & 58.73        & 73.88        & 76.49        & 72.82        & 72.80        & 91.35        & 93.09        & 90.71        & 90.77         \\ 
           &0.5      & 57.83        & 59.06        & 57.12        & 57.07        & 72.37        & 75.00        & 71.45        & 71.31        & 90.16        & 92.12        & 89.37        & 89.34         \\ 
           &0.7      & 54.34        & 55.65        & 53.81        & 53.74        & 68.74        & 71.48        & 67.58        & 67.65        & 87.68        & 90.15        & 86.69        & 87.02         \\ 
 $E_2$     &0.4      & 87.13        & 87.46        & 86.94        & 86.58        & 96.86        & 97.07        & 96.76        & 96.51        & 99.86        & 99.90        & 99.85        & 99.84         \\ 
           &0.5      & 86.05        & 86.42        & 85.84        & 85.53        & 96.45        & 96.71        & 96.33        & 96.17        & 99.84        & 99.88        & 99.84        & 99.80         \\ 
           &0.7      & 83.40        & 83.76        & 83.16        & 82.89        & 95.02        & 95.37        & 94.88        & 94.72        & 99.74        & 99.79        & 99.74        & 99.71         \\ 
 $U_g$     &0.4      & 71.08        & 74.04        & 69.76        & 72.31        & 86.13        & 89.22        & 84.76        & 87.34        & 99.56        & 99.69        & 99.53        & 99.57         \\ 
           &0.5      & 69.64        & 72.74        & 68.25        & 70.79        & 84.62        & 87.96        & 82.96        & 85.56        & 99.51        & 99.62        & 99.44        & 99.50         \\ 
           &0.7      & 66.51        & 69.75        & 64.92        & 67.97        & 81.57        & 85.46        & 79.69        & 83.20        & 99.12        & 99.36        & 98.99        & 99.10         \\ 
 \bottomrule 
 \end{longtblr}
}

It can be seen that the powers under $H_{1A}$ are generally lower than those under $H_{1B}$. This is because the discrepancy in proportions specified under $H_{1A}$ is smaller compared to that under $H_{1B}$. 
Within each power table, the Wald and LR tests are the most and second most powerful tests, respectively, for all the sample size designs and parameter configurations. Score tests and GEE tests are less powerful, but their associated powers are comparable to those of Wald and LR tests. In addition, for each given number of groups $g$ and sample design, the powers obtained by GEE tests decrease as the $R$ (under Rosner's model) or $\rho$ (under Donner's model) increases. This pattern is not observed in the other three likelihood-based tests.

Therefore, based on the simulation results, we conclude that the likelihood-based score test and the GEE test are robust under both Rosner's and Donner's model, with their resulting type I error rates are under control and powers that are generally expected, though slightly underestimated. This echoes the findings for the score test in the work of Ma and Wang K.~\cite{ma2021testing} and Ma and Wang H.~\cite{ma2022testing}, and in the meantime, recommends the GEE as an alternative approach.

\subsection{Real World Example}
\label{subsec:real-data}
Two real world examples are studied to illustrate the equal proportion test. Since there is more than one model available for analysis, we first conduct a model selection process to identify a more suitable model to describe the dataset following the goodness-of-fit test procedure~\cite{zhou2025goodness}. 

The first example consists of a subset of 214 children who were admitted because of acute otitis media with effusion (OME) and were randomized in two groups respectively treated with cefaclor and amoxicillin~\cite{mandel1982duration}. Table~\ref{tab:OME} shows the number of cured ears in 173 children at 42 days. 
\begin{table}[thpb]
    \centering
    \caption{Number of cured ears at 42 days in children treated with cefaclor and amoxicillin.}
    \label{tab:OME}
    \begin{tabular}{cccc}
    \toprule
         &\multicolumn{2}{c}{Treatment} & \\
         \cmidrule{2-3}
        \# of cured ears &Cefaclor &Amoxicillin &total \\
        \midrule
        0 &9 &7 &16 \\
        1 &7 &5 &12 \\
        2 &23 &13 &36 \\
        total &39 &25 &64 \\
        \midrule
        0 &20 &19 &39 \\
        1 &34 &36 &70 \\
        total &54 &55 &109 \\
        \bottomrule
    \end{tabular}
\end{table}

Based on the goodness-of-fit test procedure by Zhou and Ma~\cite{zhou2025goodness}, it shows that both Rosner's and Donner's models are suitable for this dataset (all p-values for goodness-of-fit test $\gtrsim0.9$), with Rosner's model yielding slightly lower AIC (274.1305 under Rosner's model vs 274.1406 under Donner's model). Therefore, we select Rosner's model to perform the equal proportion test. 
The constrained and unconstrained MLEs, along with the statistics and p-values under Rosner's model can be found in Table \ref{tab:OME:rosner}.  
\begin{table}[thpb]
    \centering
    \caption{Constrained and unconstrained MLEs along with statistics and p-values under Rosner's model for the OME dataset.}
    \label{tab:OME:rosner}
    \begin{tabular}{cccccc}
    \toprule
         &\multicolumn{2}{c}{Constrained MLEs} & &\multicolumn{2}{c}{Unconstrained MLEs}  \\
         \cmidrule{2-3}\cmidrule{5-6}  
          &Cefaclor &Amoxicillin & &Cefaclor &Amoxicillin \\
          \midrule
          $\hat{\pi}_i$ &\multicolumn{2}{c}{$\hat{\pi}_0=0.6482$} & &$\hat{\pi}_1=0.6528$ &$\hat{\pi}_2=0.6425$ \\
          $\hat{R}$ &\multicolumn{2}{c}{$\hat{R}_0=1.3182$} & &\multicolumn{2}{c}{$\hat{R}=1.3172$} \\
          $\hat{\rho}$ &\multicolumn{2}{c}{$\hat{\rho}_0=0.5862$} & &$\hat{\rho}_1=0.5964$ &$\hat{\rho}_2=0.5699$ \\
          \toprule
          &$Q_{LR}$ &$Q_W$ &$Q_S$ &$Q_{GS}$ & \\
          \midrule
          statistic &0.0394 &0.0391 &0.0395 &0.0265 & \\
          p-value &0.8426 &0.8432 &0.8424 &0.8706 & \\
          \bottomrule
    \end{tabular}
\end{table}
One can see that all the p-values from the three likelihood-based tests and from the GEE test are of similar magnitude ($\gtrsim0.8$) and much larger than the nominal level of $\alpha=0.05$. Therefore, we fail to reject the null hypothesis $H_0:~\pi_1=\pi_2$, indicating no significant difference in the proportion of the cured ears between the two treatments (cefaclor versus amoxicillin). This conclusion aligns with the findings in the original study~\cite{mandel1982duration}, which reported that `\textit{by 42 days after entry the percentage of children whose ears were without effusion or ``improved'' was equal in both treatment groups (68.9\% in the cefaclor group and 67.5\% in the amoxicillin group)}', despite overlooking the intra-subject correlation. However, by taking the intra-subject correlation into account in the likelihood-based tests and the GEE test, we observe stronger evidence for the equal proportion of cured ears across treatment groups, as reflected by very large p-values.

The second example involves the ophthalmologic study conducted in the Massachusetts Eye and Ear Infirmary between 1970 and 1979, where data were collected from an outpatient population of patients with retinitis pigmentosa (RP) and their normal relatives~\cite{berson1980risk}. Patients were classified into four types of genetic groups: (i) autosomal dominant RP (DOM), (ii) autosomal recessive RP (AR), (iii) sex-linked RP (SL), and (iv) isolate RP (ISO).
A subset of 218 patients aged 20 - 39 was originally analyzed by Rosner using the constant $R$ model~\cite{Rosner_1982}, where an eye was considered affected if the best corrected Snellen visual acuity (VA) was 20/50 or worse, and normal if VA was 20/40 or better. 
Table~\ref{tab:RP} presents the distribution of the number of affected eyes for 216 patients in the four genetic groups who had complete information for VA on both eyes.
This example can be considered as a special case in the combined data framework where unilateral observations are absent. 
\begin{table}[thpb]
    \centering
    \caption{Number of affected eyes for patients in four genetic groups.}
    \label{tab:RP}
    \begin{tabular}{cccccc}
        \toprule
         &\multicolumn{4}{c}{Genetic Type} & \\
         \cmidrule{2-5}
        \# of affected eyes &DOM &AR &SL &ISO &total \\
        \midrule
        0 &15 &7 &3 &67 &92 \\
        1 &6 &5 &2 &24 &37 \\
        2 &7 &9 &14 &57 &87 \\
        total &28 &21 &19 &148 &216 \\
        \bottomrule
    \end{tabular}
\end{table}

Using the goodness-of-fit test results presented in~\cite{zhou2025goodness}, one can see that Donner's model is much superior to Rosner's model as reflected by one magnitude larger p-values and smaller AIC (443.7967 under Donner's model vs 449.9490 under Rosner's model). Therefore, we select Donner's model to perform the equal proportion test.  
The constrained and unconstrained MLEs, along with the statistics and p-values under Donner's model, are presented in Table \ref{tab:RP:donner}. As can be seen, all the p-values are smaller than $0.05$, with the p-values from the score and GEE tests being close to each other. Therefore, we reject the null hypothesis $H_0:~\pi_1=\pi_2=\pi_3=\pi_4$ and conclude that there is an overall difference in proportions of the affected eyes across the four genetic groups. This finding is consistent with the original paper by Rosner \cite{Rosner_1982}, which also noted that the overall difference was completely attributed to the differences between the SL group and each of the other three groups. 
\begin{table}[thpb]
    \centering
    \caption{Constrained and unconstrained MLEs along with statistics and p-values under Donner's model for the retinitis pigmentosa dataset.}
    \label{tab:RP:donner}
    \begin{tabular}{ccccc}
      \toprule
      &\multicolumn{4}{c}{Constrained MLEs} \\
      \midrule
      &DOM &AR &SL &ISO \\
      \cmidrule{2-5}
      $\hat{\pi}_i$ &\multicolumn{4}{c}{$\hat{\pi}_0=0.4884$} \\
      $\hat{\rho}$ &\multicolumn{4}{c}{$\hat{\rho}_0=0.6572$} \\
      \midrule
      &\multicolumn{4}{c}{Unconstrained MLEs} \\
      \midrule
      &DOM &AR &SL &ISO \\
      \cmidrule{2-5}
      $\hat{\pi}_i$ &$\hat{\pi}_1=0.3625$ &$\hat{\pi}_2=0.5455$ &$\hat{\pi}_3=0.7926$ &$\hat{\pi}_4=0.4658$ \\
      $\hat{\rho}$ &\multicolumn{4}{c}{$\hat{\rho}=0.6416$} \\
      \bottomrule
      &$Q_{LR}$ &$Q_W$ &$Q_S$ &$Q_{GS}$ \\
      \midrule
      statistic &12.0385 &16.3267 &11.3158 &10.6890 \\
      p-value &0.0073 &0.0010 &0.0101 &0.0135 \\
      \bottomrule
    \end{tabular}
\end{table}

\section{Discussion and Conclusions}
\label{sec:conclusions}
In this study, we consider the GEE approach as an alternative to three likelihood-based methods (likelihood ratio, Wald-type, and score statistics) for testing homogeneity of proportions across $g\ge2$ groups for the combined unilateral and bilateral data. We revisit the likelihood-based methods under two statistical models: i) Rosner's constant $R$ model and ii) Donner's equal correlation $\rho$ model, accounting for intra-subject correlations in the bilateral portion, and introduce the GEE method along with its implementation in \texttt{SAS GENMOD} procedure for the combined organ data. 

Monte Carlo simulations evaluate the empirical type I error rates and powers of each method under various sample sizes and parameter configurations. Results show that the GEE and score tests exhibit comparable performance, outperforming the likelihood ratio and Wald-type tests in controlling type I error. All tests demonstrate similar power, though the likelihood ratio and Wald-type tests are slightly inflated. These findings confirm that the GEE test performs at least as well as the score test and revalidate previous results reported by Ma and others~\cite{ma2021testing,ma2022testing}. Importantly, the GEE method offers additional flexibility, allowing incorporation of continuous covariates and more complex model structures, which provides an advantage over the score test.

Applications to two real datasets from otolaryngologic and ophthalmologic studies illustrate these methods. Goodness-of-fit tests guide model selection for the likelihood-based methods, showing that Rosner's model is preferred for the OME dataset, while Donner's model better fits the retinitis pigmentosa dataset. Inference from both the GEE method and the likelihood-based methods after model selection are consistent with the original study results.

In conclusion, we recommend the GEE and score tests for homogeneity testing for the combined unilateral and bilateral data. Both methods provide robust type I error control and strong power across different scenarios. For extended analyses involving continuous covariates or multiple explanatory variables, the GEE approach is preferred. The score test is computationally efficient, particularly for large samples or numerous groups, and is well suited for exact tests, such as Fisher's exact or permutation tests, when data are sparse, which is less straightforward with standard software implementation of GEE.

\vspace{2em}
\declare{Author contributions}{The authors confirm contribution to the paper as follows: study conception and design: Ma C-X; analysis and interpretation of results: Zhou J, Ma C-X; draft manuscript preparation: Zhou J. All authors reviewed the results and approved the final version of the manuscript.}

\declare{Conflict of interest}{The authors declare that they have no conflict of interest.}

\declare{Data availability}{The data presented in this study are openly available in references~\cite{mandel1982duration,Rosner_1982}.}

\bibliographystyle{unsrt}
\bibliography{correlateddata}

\end{document}